\documentclass[twocolumn,aps,apl,floatfix,reprint]{revtex4-1}

\usepackage{graphicx}
\usepackage{epstopdf}
\usepackage{amsmath}
\usepackage{latexsym}
\usepackage{amsmath}
\usepackage{esdiff}
\usepackage{xcolor}

\newcommand{\beq}{\begin{equation}}
\newcommand{\eeq}{\end{equation}}

\begin{document}

\title{Atomistic study of an ideal metal/thermoelectric contact: the full-Heusler/half-Heusler interface}

\author{Catalin D. Spataru}
\thanks{cdspata@sandia.gov}
\author{Yuping He}
\author{Fran\c{c}ois L\'{e}onard}
\affiliation{Sandia National Laboratories, Livermore, California 94551, USA}

\begin{abstract}
Half-Heusler alloys such as the (Zr,Hf)NiSn intermetallic compounds are important thermoelectric materials for converting waste heat into electricity. Reduced electrical resistivity at the hot interface between the half-Heusler material and a metal contact is critical for device performance, however this has yet to be achieved in practice. 
Recent experimental work suggests that a coherent interface between half-Heusler and full-Heusler compounds can form due to diffusion of transition metal atoms into the vacant sublattice of the half-Heusler lattice. We study theoretically the structural and electronic properties of such an interface using a first-principles based approach that combines {\it ab initio} calculations with macroscopic modeling. We find that the prototypical interface HfNi$_2$Sn/HfNiSn provides very low contact resistivity and almost ohmic behavior over a wide range of temperatures and doping levels. Given the potential of these interfaces to remain stable over a wide range of temperatures, our study suggests that full-Heuslers might provide nearly ideal electrical contacts to half-Heuslers that can be harnessed for efficient thermoelectric generator devices.
\end{abstract}

\maketitle

Half-Heusler (HH) alloys are promising candidate materials for thermoelectric (TE) applications in the medium to high temperature range, due to their good thermal, mechanical and chemical stability, as well as good TE figure of merit ZT \cite{Kanatzidis,Snyder,Huang}. HH materials possess intrinsically high charge mobility that enables large power factors \cite{Chen18}. Mass fluctuation via chemical substitution as well as doping can be used to engineer low lattice thermal conductivity \cite{Shen,Poon2}. 
Within the family of HH alloys, (Zr,Hf)NiSn compounds have shown the largest ZT values $\sim1$ achieved at the optimal temperature of $600 ^{\circ}$C \cite{Huang,Zillmann}. Materials in this class are relatively low-cost \cite{LeBlanc}, easy to synthesize and they represent an ideal material candidate for TE power generators (TEG) applications \cite{Rama,Priya}. 

An important obstacle that needs to be overcome in order to realize efficient TEG operating at high temperature is achieving reduced electrical contact resistivity between the TE material and the metal contact \cite{Ngan}. For optimal device performance, the contact resistance at the TE material-metal interface should be a small fraction of the total TE module resistance, with typical desired values lower than $10^{-6}\Omega$cm$^2$ for compact TE modules with length $\sim 1$ mm and bulk TE intrinsic resistivity  $\sim10^{-3}\Omega$cm \cite{Rama}. This poses serious challenges because with increasing temperature a number of complex phenomena such as oxidation, reactivity and inter-diffusion get enhanced and may impact adversely the contact properties \cite{Ngan}. 
So far, little attention has been paid towards understanding the properties of metal contacts to HH alloys, and experimental and theoretical work is needed to understand the nature of the interfacial structure and chemical composition of HH-metal interfaces.

In ambient atmosphere (Zr,Hf)NiSn compounds are resistant against oxidation up to temperatures $\sim600 ^{\circ}$C. Unfortunately, most metal contacts  degrade quickly in air at such operating temperatures. An exception is Au which however suffers from poor adhesion to HH. Practical approaches to oxidation protection include module level encapsulation, for example in inert atmosphere, or using oxidation-resistant coatings. These approaches however do not prevent reactivity or inter-diffusion processes from taking place at the metal-TE material interface. 
A recent experimental study \cite{Ngan} of Ag contacts to HH alloys indicates the presence of a $\mu$m-scale reaction zone rich in intermetallic phases such as Ag$_3$Sn that forms at high temperature. While the electronic properties of the interfaces between these intermetallic phases and HH materials have not been quantified,  it has been suggested that electron injection barriers ({\it e.g.} Schottky) may be small enough to be overcome at high temperature. Unfortunately these contacts develop at high temperature large microgaps and the contact resistivity is high with typical values larger than $\sim10^{-4}\Omega$cm$^2$.
 
A potentially good metal contact to (Zr,Hf)NiSn compounds that can offer good thermal stability and high mechanical strength is Ni \cite{Ren,Snyder}. Ni provides better match than Ag for the HH coefficient of thermal expansion (CTE) so the Ni contacts may be less prone to formation of microvoids. The nature of reactivity/inter-diffusion between Ni and the HH material is not fully understood. However, recent experiments \cite{Makongo_JSSC,Makongo_JACS} show that HH alloys such as Zr$_{0.75}$Hf$_{0.25}$NiSn react with elemental Ni at high temperature to form HH/full Heusler (FH) composites with coherent phase boundaries. The composites arise from solid-state Ni diffusion into the vacant sites of the HH structure, facilitated by the similarity in the crystal structures between HH and FH. Indeed, the Heusler crystal structure consists of four interpenetrating fcc sublattices. In the HH phase one of these sublattices is vacant \cite{Poon}. In the specific case of (Zr,Hf)NiSn compounds, filling the Ni vacant sublattice results in the full-Heusler (FH) compound (Zr,Hf)Ni$_2$Sn. 

The aforementioned experimental findings suggest that it is possible that upon deposition of Ni on a (Zr,Hf)NiSn substrate, with increasing temperature, the Ni overlayer will react to fill the unoccupied fcc sublattice of the HH, thereby forming the FH structure (Zr,Hf)Ni$_2$Sn.
Recent first-principles calculations \cite{Page} of the (Zr,Hf)Ni$_{1+x}$Sn phase diagram indicate that a large miscibility gap separates HH from FH. In particular the solubility limit of excess Ni in the HH structure is found negligible at room temperature and increases to only $1\%$ at 1450 K. This suggests that the formation of a transition layer such as (Zr,Hf)Ni$_{1+x}$Sn may not be favorable at the interface between HH and FH.
One expects that inter-diffusion of Ni in HH forms a self-limiting FH layer providing a coherent, stable interface to HH with contact properties that could be harnessed for TEG applications. We explore this scenario and present a theoretical study of the electronic properties of HH/FH interfaces. In particular we are interested in the fundamental limit of the electrical contact resistivity at these interfaces. 
To this end we employ a previously developed atomistic approach\cite{AEM18,PRM18} that combines first-principles simulations with macroscopic modeling. The approach uses: i) {\it ab initio} calculations based on Density Functional Theory (DFT) \cite{DFT} to obtain the electronic properties near the interface. ii) macroscopic modeling to extend these results away from the interface and for a variety of doping levels and temperatures and ultimately estimate the contact resistivity.

We begin by investigating from first-principles the basic properties of the HH and FH systems in bulk form. In a previous study \cite{PRM18} we calculated the low-temperature properties of Ag contacts to several HH alloys including compounds from the family (Zr,Hf)NiSn. In particular we found that bulk HH properties important for band alignment at the metal interface (such as the electron affinity or the band gap) do not depend significantly on the alloy composition or doping. Taking advantage of this we focus on the undoped, end point HH alloy, namely the ternary intermetallic compound HfNiSn. Correspondingly we consider HfNi$_2$Sn as the FH counterpart. This particular choice allows us to construct interfacial structures via supercells with minimal lateral size ({\it i.e.} along directions parallel to the interface) while still capturing the phenomena relevant to electron transport across the interface. 

The DFT calculations are performed using the VASP code \cite{Kresse} in conjunction with projector augmented wave pseudo-potentials \cite{PAW}. We treat the exchange and correlation terms within the generalized gradient approximation (GGA) with the PW91 parametrization \cite{GGA}. The valence electron configurations of 5p$^6$6s$^2$5d$^2$, 4s$^2$3d$^8$ and 5s$^2$5p$^2$ were used for Hf, Ni and Sn atoms, respectively. A kinetic energy cutoff of 450 eV was used for the expansion of the plane-wave basis set for all systems.
 
\begin{figure}
\vspace{-0.0cm}
\begin{center}
\includegraphics[trim=100 50 250 50,clip,width=\columnwidth]{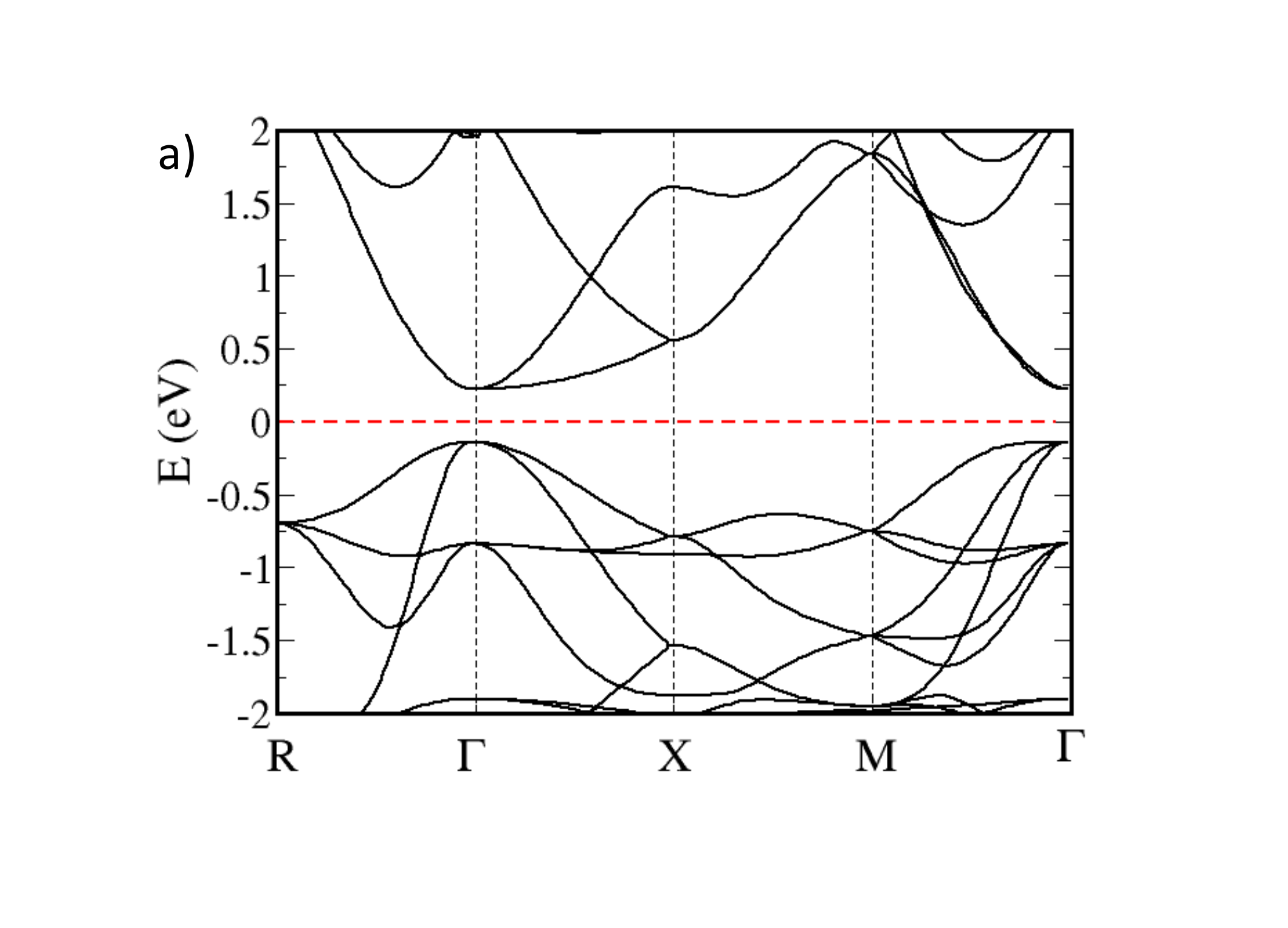}
\includegraphics[trim=100 50 250 50,clip,width=\columnwidth]{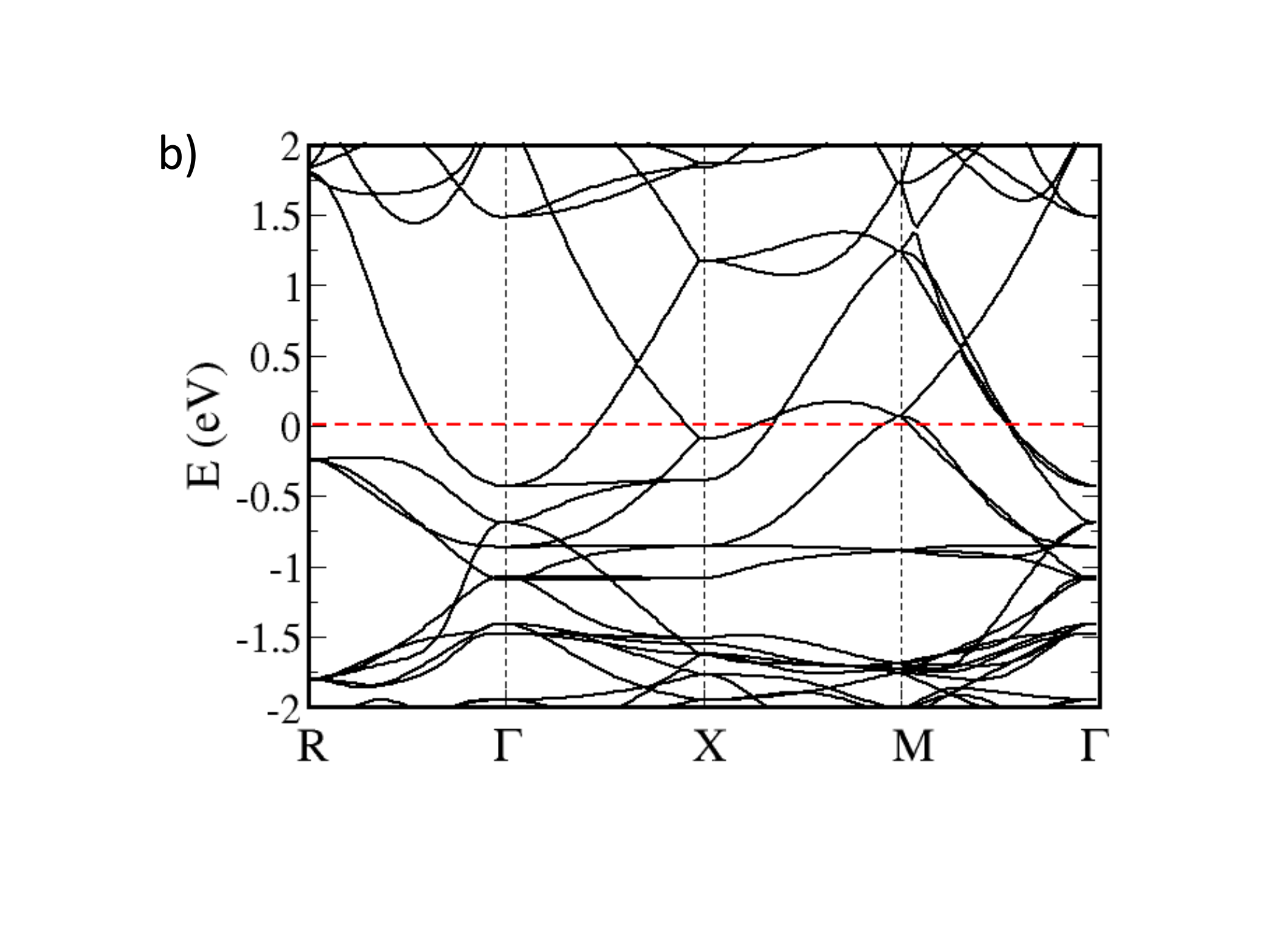}
\includegraphics[trim=-100 0 -60 0,clip,width=\columnwidth]{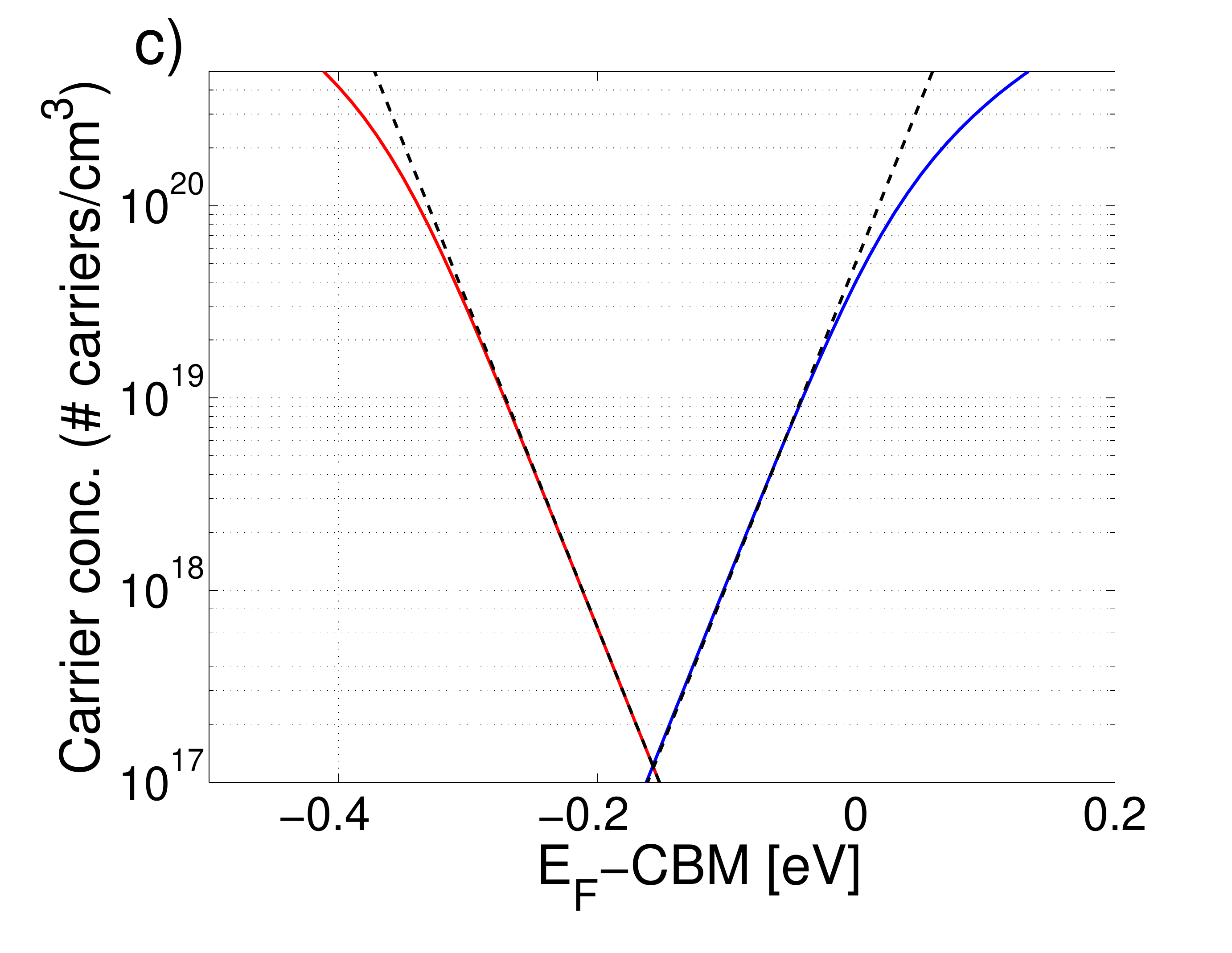}
\end{center}
\caption{Electronic bandstructure of: a) the HH HfNiSn, b) the FH HfNi$_2$Sn, calculated along several high symmetry directions in the Brillouin zone. c) Carrier concentration (red line for holes, blue line for electrons) as function of Fermi level for the HH HfNiSn.}
\vspace{-0.0cm}
\label{bandstr_HH_FH}
\end{figure}

The optimized unit cell of HfNiSn is cubic with length $6.12$ \AA\ and the system is a semiconductor with a calculated electronic band gap of $0.34$ eV. The electronic band structure of HfNiSn is shown in Fig.~\ref{bandstr_HH_FH}a) along several high symmetry directions in the Brillouin zone. By contrast HfNi$_2$Sn is metallic, as indicated in Fig.~\ref{bandstr_HH_FH}b). The filling of the Ni sublattice increases the HfNi$_2$Sn unit cell by $\sim$2\%  compared to HfNiSn, in good agreement with previous experimental and theory results \cite{Rabe}. Based on existing literature data on the CTE of the HH, and those of related HH and FH compounds \cite{Hermet,Jung} we expect the lattice mismatch between the FH and HH to increase by only $0.15\%$ as the temperature varies from ambient to the optimal operating conditions \cite{CTE}. 

Another bulk property of relevance for interfacial electron transport phenomena is the
 carrier concentration in the semiconductor as function of the Fermi level $E_F$. Figure \ref{bandstr_HH_FH}c) shows the concentration of holes and electrons estimated at room temperature via standard expressions \cite{PRM18} based on DFT eigenvalues calculated for a dense set (36x36x36) of Monkhorst-Pack k-points in the Brillouin zone \cite{Monk}. N-type doping can be achieved in (Zr,Hf)NiSn compounds via Sn substitution with Sb. Doping levels that optimize the TE properties are typically of the order of $10^{20}$ e/cm$^3$. According to Fig.~\ref{bandstr_HH_FH}c) this corresponds to $E_F$ situated several tens of meV above the conduction band minimum (CBM). We also note that in the limit of zero doping (intrinsic semiconductor) $E_F$ lies close to the mid-gap.
 
\begin{figure}
\vspace{-0.0cm}
\begin{center}
\includegraphics[width=\columnwidth]{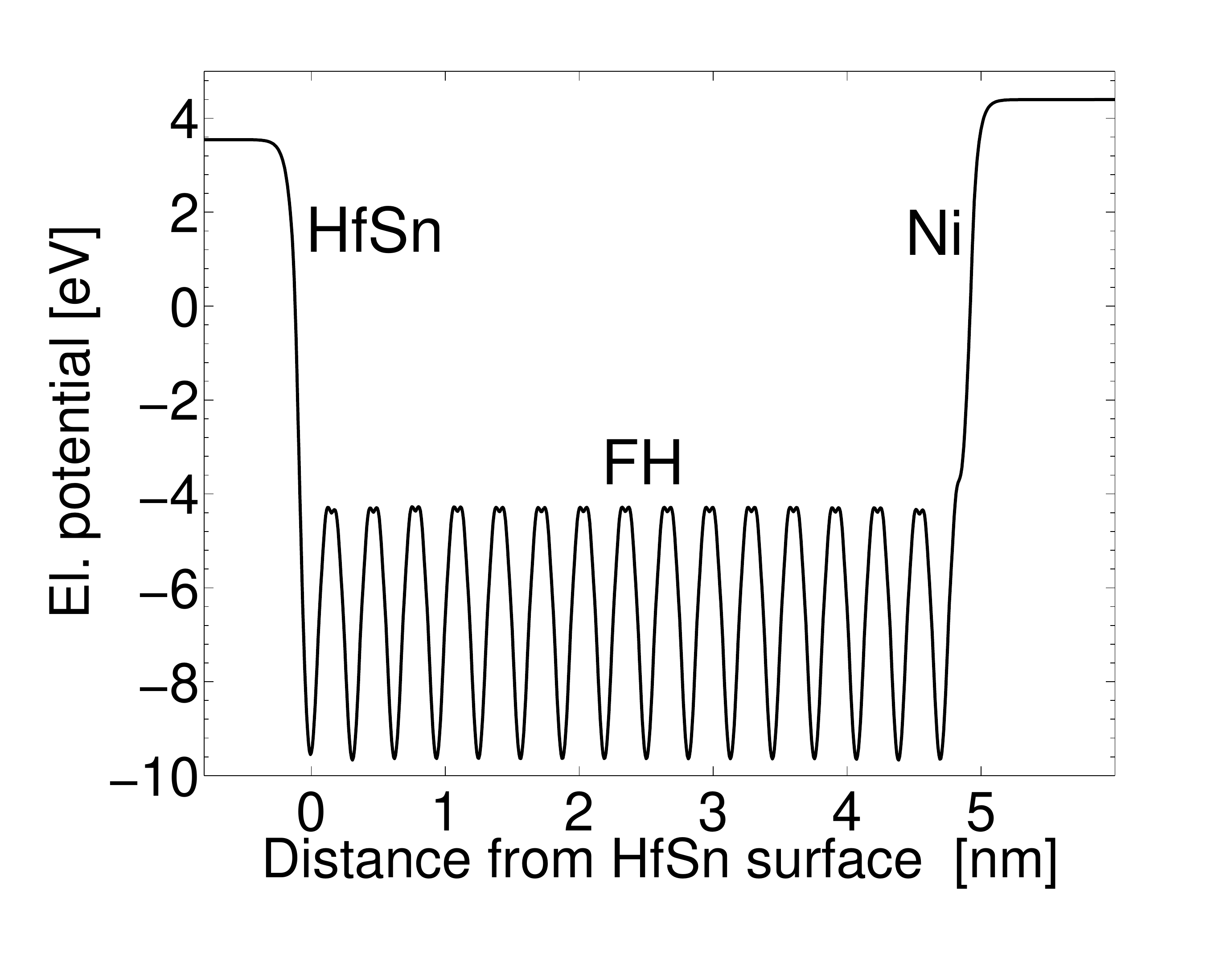}
\end{center}
\caption{Planar averaged electrostatic potential for an isolated FH  HfNi$_2$Sn slab of width $\approx5$ nm. The Fermi level is at 0 eV.}
\vspace{-0.0cm}
\label{Electr_pot}
\end{figure}

\begin{figure}
\vspace{-0.0cm}
\begin{center}
\includegraphics[trim=200 200 300 150,clip,width=\columnwidth]{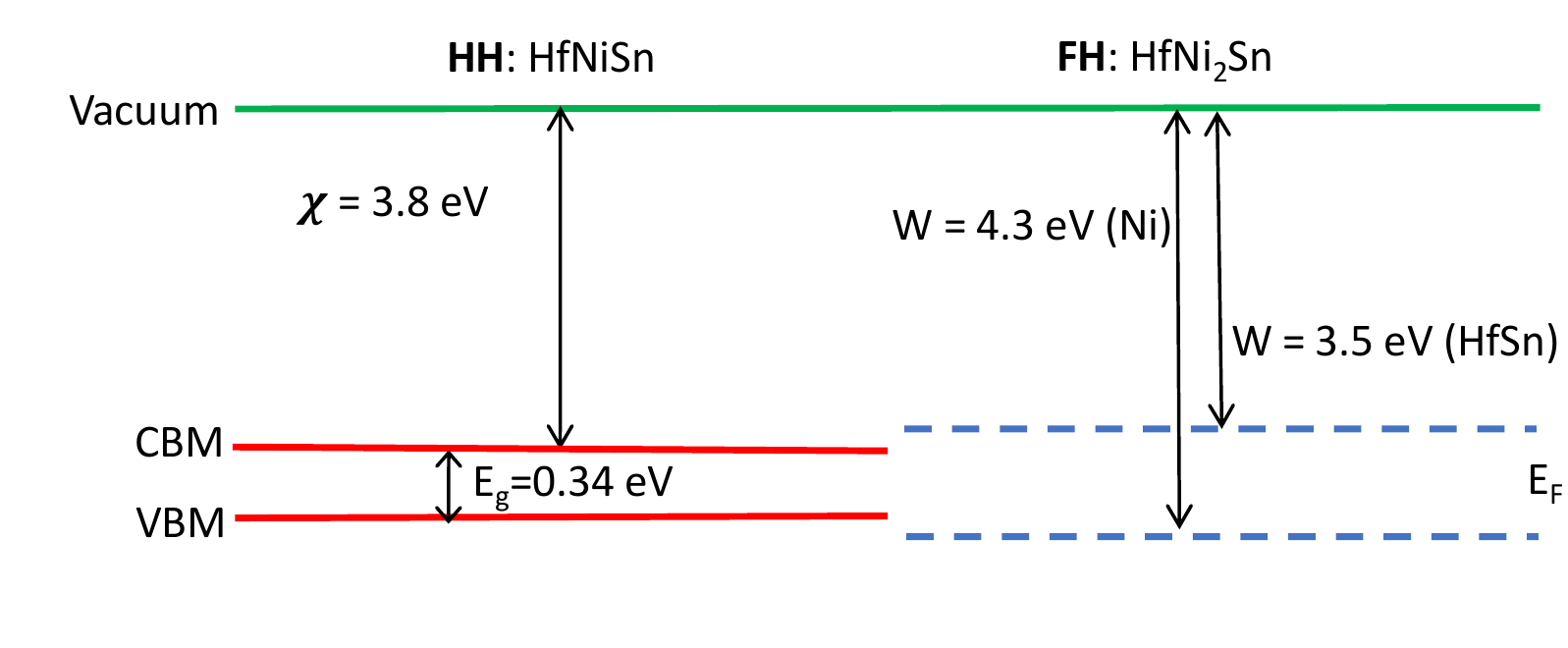}
\end{center}
\caption{Schematic bare band alignment at the interface between HH HfNiSn and FH  HfNi$_2$Sn, based on the calculated work functions (W) of the two systems.}
\vspace{-0.0cm}
\label{bare_alignement}
\end{figure}

\begin{figure}
\vspace{-0.0cm}
\begin{center}
\includegraphics[trim=50 200 450 50,clip,width=\columnwidth]{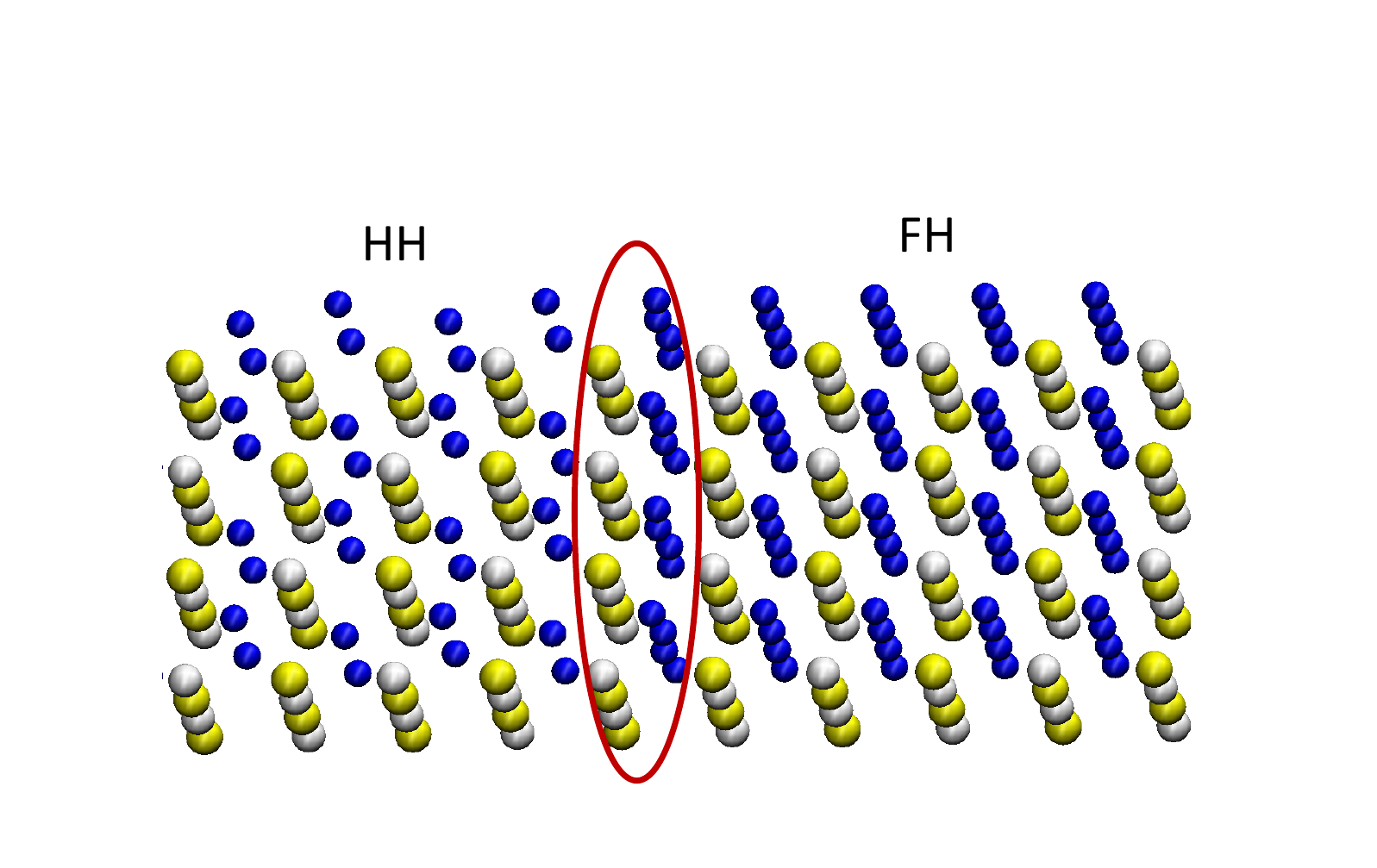}
\end{center}
\caption{Optimized atomic structure of the HH/FH interface. The oval indicates two atomic layers (HfSn -atoms colored in white and yellow- or Ni -blue colored atoms-) that could be thought as terminating the FH side contacted by HH.}
\vspace{-0.0cm}
\label{structure}
\end{figure}

The conventional Schottky-Mott rule is often used for a zero-order estimate of the Schottky barrier height at a metal-semiconductor interface. This approach uses the work function $W$ of the metal and 
the semiconductor electron affinity $\chi$ to obtain the bare electronic alignment in the absence of atomistic effects at the interface. It is instructive to use this rule for the FH/HH contact. For that, we obtain $\chi$ for the HH and $W$ for the FH from the vacuum electrostatic potential level w.r.t. $E_F$ calculated for isolated HH and FH slabs. We consider slab surfaces perpendicular to the [001] direction and terminated with either HfSn or Ni layers. We find that $\chi=3.8$ eV irrespective of the HH surface termination. The calculated planar averaged electrostatic potential for the FH is shown in Fig.~\ref{Electr_pot} (we use dipole corrections to avoid interactions between periodically repeated slab images). We note that because the atomic structure of the HH and FH are so similar, there is an ambiguity in choosing the surface termination for each at the interface. Since $W$ is sensitive to the surface termination with values differing by more than $0.7$ eV between the two cases,
this results in an ambiguity as to which $W$ to use in the Schottky-Mott rule (see also Fig.~\ref{structure}): assuming that FH terminates with Ni at the interface, the bare alignment shows (see Fig.~\ref{bare_alignement}) the metal Fermi level $E_F^{FH}$ within the valence band of the HH; by contrast $E_F^{FH}$ lies within the conduction band for the HfSn termination. This implies that the Schottky-Mott rule cannot clearly establish whether the contact is Schottky or ohmic. 

An atomistic approach that includes interface interaction effects and does not suffer from the above mentioned ambiguity is necessary in order to disentangle the nature of the contact. The interface chemistry can alter the bare band alignment via formation of interface dipoles originating from complex charge redistribution phenomena related to atomic rearrangement, charge transfer into surface states, or electronegativity equalization between contacted atomic layers \cite{Tung14}. These important effects are captured by {\it ab initio} simulations of the full interface.

We simulate the full FH/HH interfacial structure with the interface perpendicular to the [001] direction. A slab supercell is constructed by joining HH and FH slabs with different surface terminations at the interface. The surface unit cells for the HH and the FH are well matched with less than $\sim1\%$ tensile/compressive lateral strain \cite{Stokbro}. Convergence with respect to k-point sampling is achieved using 12x12x1 $\Gamma$-centered k-point grids.
 The interfacial structure is relaxed until the atomic forces are smaller than $0.01$ eV/\AA{}. The optimized structure is shown in Fig.~\ref{structure}. The slight relaxation of the Ni-layer terminating the FH surface is well captured with single lateral replica of the unit cell. Along the [001] direction the HH unit cell is replicated 32 times with the HH slab width reaching $\sim20$ nm. The width of the FH slab is set to $\sim10$ nm. A large HH slab width is needed in order for interface/surface effects to fade away far from the interface \cite{short_slab}. 
 
The impact of the FH metal contact on the HH electronic band structure can be seen from the spectral function $A$ of the full interfacial structure projected on the HH side. We define $A$ projected on a slab $j$ of width $w$ located at a distance (w.r.t. the interface) ranging from $d_j-w$ to $d_j$ as \cite{AEM18,bi2se3}:
$A_j ({\bf k},E )=\sum_{n,i}^{i\in j}w_{n{\bf k}}^i \delta(E-\epsilon_{n{\bf k}})$
where $w_{n{\bf k}}^i$ is the site-projected character of the wave function of an electron characterized by band index $n$, wavevector ${\bf k}$, energy $\epsilon$, and atomic site index $i$. The $\delta$ function is approximated by a Lorentzian function of width equal to $10$ meV. 

\begin{figure}
\vspace{-0.0cm}
\begin{center}
\includegraphics[trim=0 50 0 50,clip,width=\columnwidth]{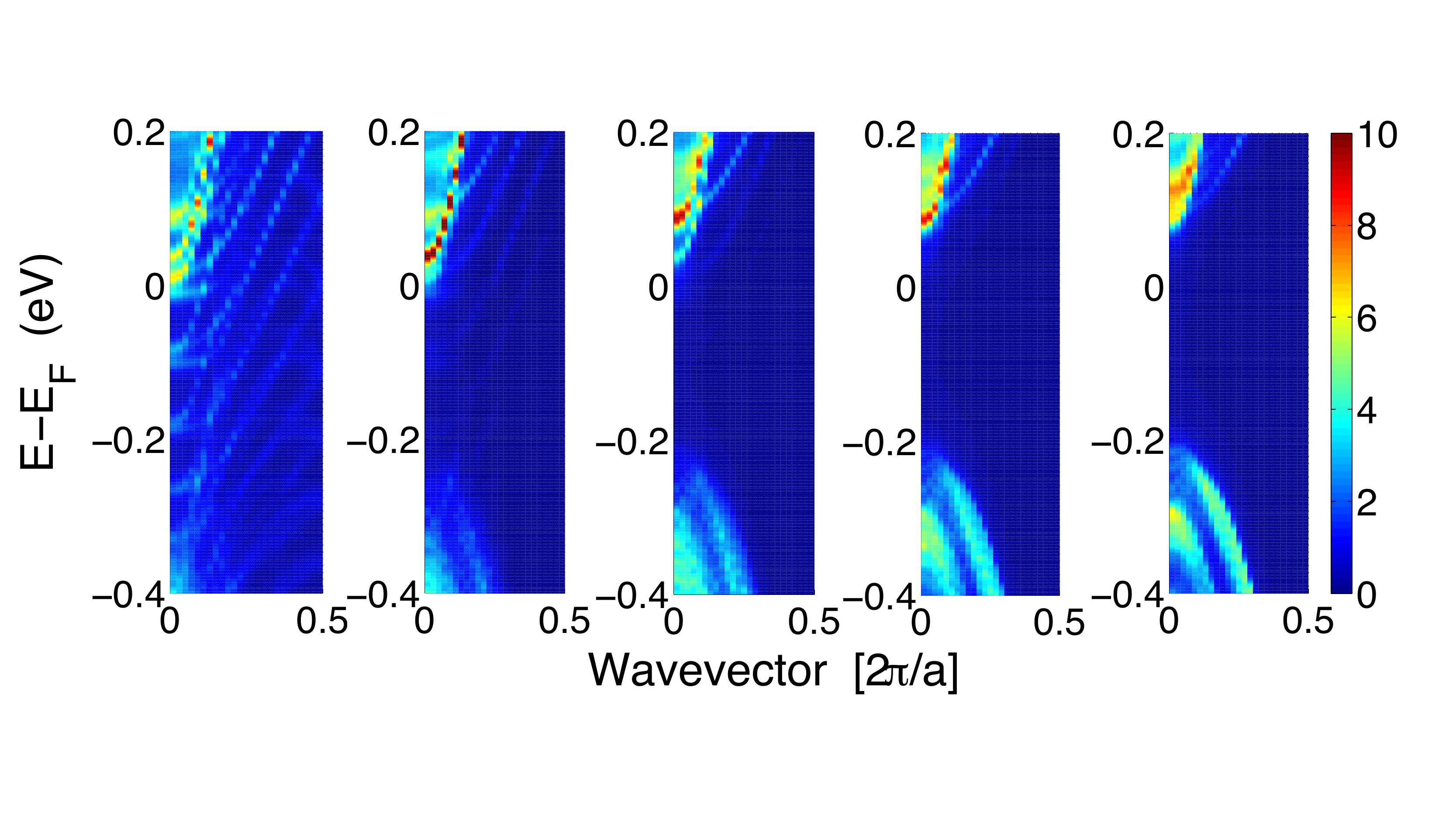}
\includegraphics[trim=0 50 10 50,clip,width=\columnwidth]{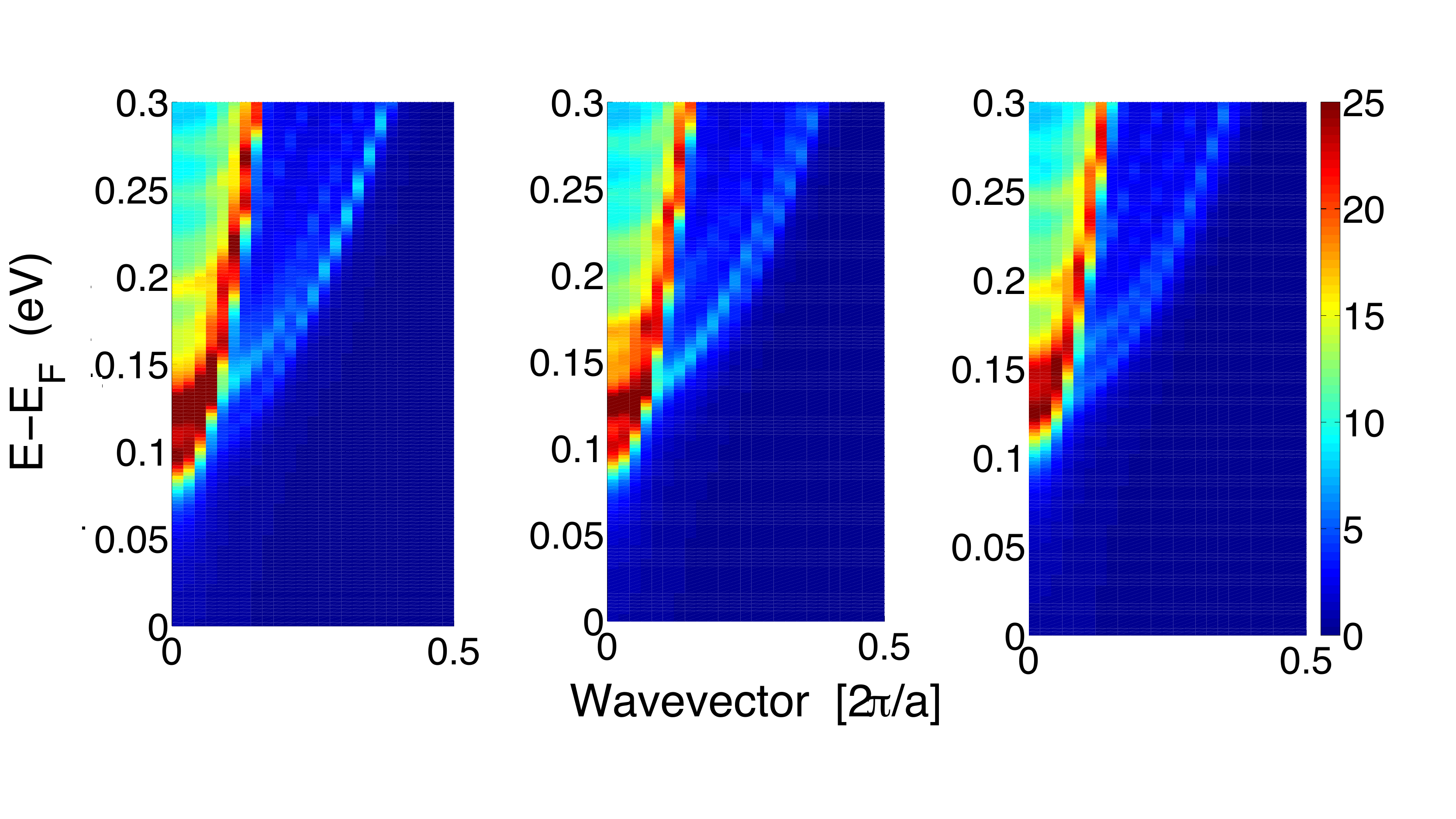}
\end{center}
\caption{Projected electronic bandstructure of the FH/HH interface. a) The spectral function is projected inside the HH on a slab $j$ of width $\sim0.6$ nm and shown for several distances $d_j$ between the interface and the side of the slab farthest from the interface (i.e. $d_j$= 0.6 nm, 1.2 nm, 1.8 nm, 2.4 nm and 3.0 nm) b) Similar to a) for slabs of width $\sim2$ nm and $d_j$ spanning a range from $\sim5$ nm to $\sim9$ nm.}
\vspace{-0.0cm}
\label{proj_bands}
\end{figure}

The {\it ab initio} calculated projected bandstructure \cite{AEM18,bi2se3} is shown projected near the interface in Fig.~\ref{proj_bands}a) for slabs with $w\sim0.6$ nm. The HH band gap can be seen for $d_j>0.6$ nm, while right near the interface one can see the fingerprint of metallic FH states extending inside the band gap. CBM can be clearly identified, and its relative position w.r.t. the Fermi level in the metal (the reference zero energy) is seen to rise slowly away from a value close to zero at the interface to positive values as one moves away from interface. The slowly varying CBM band profile is due to the fact that the HH is intrinsically undoped which yields a band bending extending over several tens of nm. This can be seen from Fig.~\ref{proj_bands}b) which shows a similar projection but farther inside the HH, using larger slabs with $w\sim2$ nm. Deep inside the HH at distances $\sim10$ nm away from interface the anchoring of CBM is in agreement with bulk results, approaching a value close to the mid-gap. 

\begin{figure}
\vspace{-0.0cm}
\begin{center}
\includegraphics[width=\columnwidth]{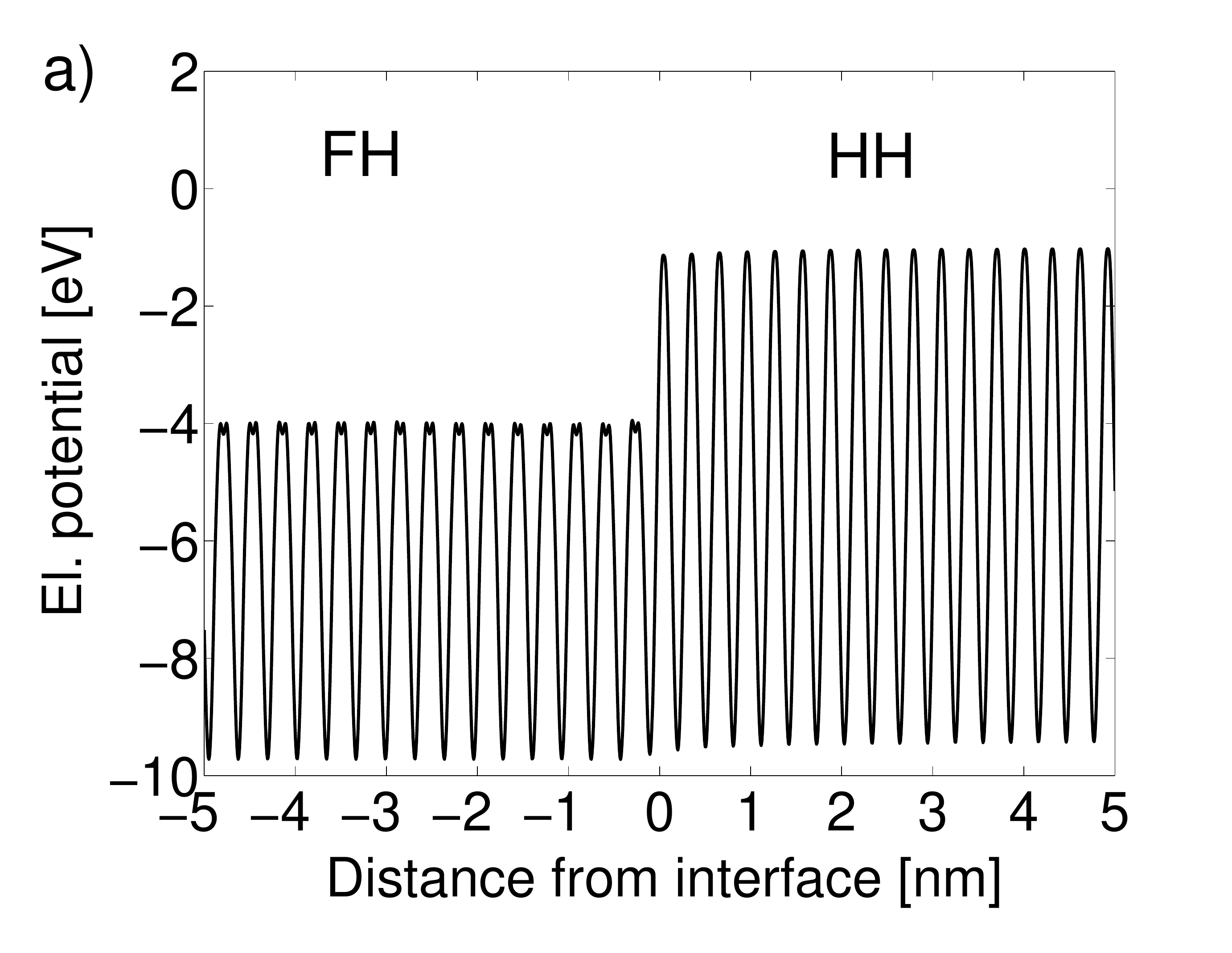}
\includegraphics[width=\columnwidth]{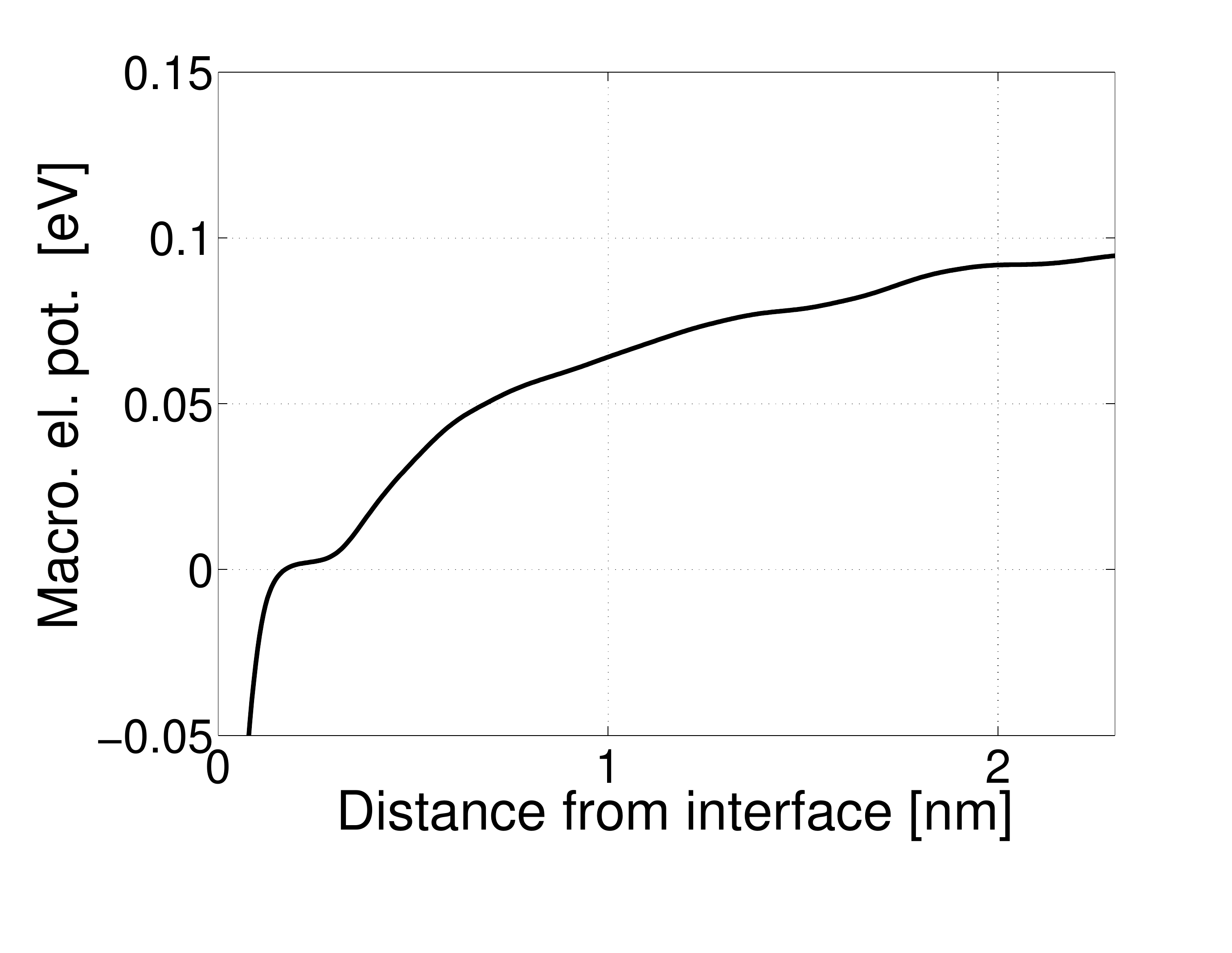}
\end{center}
\caption{a) Planar-averaged electrostatic potential $V$ of the full FH/HF interfacial structure. Fermi level is at $0$ eV. b) Macroscopic electrostatic potential profile $V_{macro}$ obtained from $V$ via the smoothing procedure described by Eq.~\ref{macro_avg}.}
\vspace{-0.0cm}
\label{el_pot}
\end{figure}

The HH CBM band profile indicates an upward shift as one moves away from the interface. We would like to reproduce this behavior using a different approach that uses the electrostatic potential as a rigid-shift of the bandstructure. Figure \ref{el_pot}a) shows the {\it ab initio} calculated planar-averaged electrostatic potential $V$ of the FH/HH interfacial structure (along directions parallel to the interface). It contains contributions from all charges in the system, including inner-shell electrons and ions that give rise to the large oscillations in the potential profile. Following Ref.~\cite{Dandrea} one defines a smooth, macroscopic band bending potential $V_{macro}$ via a a double-average integral of $V$ along the transport direction $z$ perpendicular to the HH/FH interface:
\begin{equation}
V_{macro}(z)=\frac{1}{l_{FH} l_{HH}} \int_{-l_{FH}/2}^{l_{FH}/2}dz'\int_{-l_{HH}/2}^{l_{HH}/2}dz"V(z+z'+z").
\label{macro_avg}
\end{equation}
Since the lattice constants of the HH and FH are similar we set the integration lengths appearing in 
Eq.~\ref{macro_avg} to be equal: $l_{FH}=l_{HH}=l$. A length scale $l$ approximately equal to half the lattice constant of either the bulk FH or bulk HH suffices to obtain a smooth band bending potential, as demonstrated in Fig.~\ref{proj_bands}.  
Far away from interface the potential is anchored according to the CBM position seen in Fig.~\ref{el_pot}b), {\it i.e.} about half-band gap above the Fermi level. Very good agreement between the calculated CBM band profile and the CBM level position identifiable in Fig.~\ref{proj_bands} is obtained for a wide range of distances away from interface.

Both the calculated band bending as well as the projected bandstructure indicate that near the interface the CBM aligns very close to the Fermi level. 
Far from interface the alignment is dictated by the requirement that the semiconductor is charge-neutral: depending on the doping level, the alignment of CBM w.r.t. the Fermi level is such that the free carrier concentration equals the ionic charge due to dopants. According to Fig.~\ref{bandstr_HH_FH}c), one expects that far from the interface the band bending profile will vary by $\sim150$ meV  when the doping level changes from $0$ to $10^{20}$ el/cm$^3$. 
Thus the FH contact to n-doped HH material is expected  to be ohmic for large enough doping.

Having obtained from first-principles the band bending for the undoped semiconductor case we would like to extend the results to non-zero n-type doping levels. This is done in several steps \cite{AEM18,PRM18,Codes}:

i) We obtain the macroscopic total charge density $\rho_{macro}$ (including free carriers, valence and ionic charge) from the one-dimensional Poisson equation satisfied by $V_{macro}$:
\beq
\frac{\partial^2 V_{macro}}{\partial z^2}=-\rho_{macro}(z)
\label{eq_i}
\eeq
ii) We rewrite the Poisson equation taking into account the screening response of the valence electrons via the dielectric constant $\epsilon$:
\beq
\frac{\partial^2 V_{macro}}{\partial z^2}=-\rho_{0}(z)+e\frac{n(z)-p(z)}{\epsilon}
\label{eq_ii}
\eeq
where $n$/$p$ is the free carrier density of electrons/holes which depends on the Fermi level position with respect to CBM (via the rigid shift set by $V_{macro}$) as indicated in Fig.~\ref{bandstr_HH_FH}c). $\rho_0$ represents the new, localized charge induced by the FH-HH interaction. We use $\epsilon=24$ as obtained in other first principle studies \cite{Vanderbilt} and employ an effective mass fitting (indicated by the dashed lines in  Fig.~\ref{bandstr_HH_FH}c)) of the carrier concentration near CBM and VBM using electron and hole effective masses $m_e = 1.54$ m$_0$ and $m_h = 0.77$ m$_0$ together with a band degeneracy factor of 3 at CBM and VBM. The HH local band gap renormalization near the interface (see Fig. \ref{proj_bands}a)) does not appear significant enough to warrant the consideration of a correspondingly position-dependent effective mass \cite{Aldegunde}.

iii) We finally extend the band bending results to the doped case characterized by uniform background doping $N_d$. We use the fact that $\rho_0$ is insensitive to $N_d$ \cite{AEM18} and find the new potential $\tilde{V}_{macro}$ and free carrier concentration $\tilde{n},\tilde{p}$ by solving self-consistently the Poisson equation:
\beq
\frac{\partial^2 \tilde{V}_{macro}}{\partial z^2}=-\rho_{0}(z)+e\frac{\tilde{n}(z)-\tilde{p}(z)+N_d}{\epsilon}
\label{eq_iii}
\eeq
with charge-neutrality conditions imposed far away from interface, namely $n(z) - p(z) + N_d = 0$ as $z\rightarrow \infty$.

\begin{figure}
\vspace{-0.0cm}
\begin{center}
\includegraphics[width=\columnwidth]{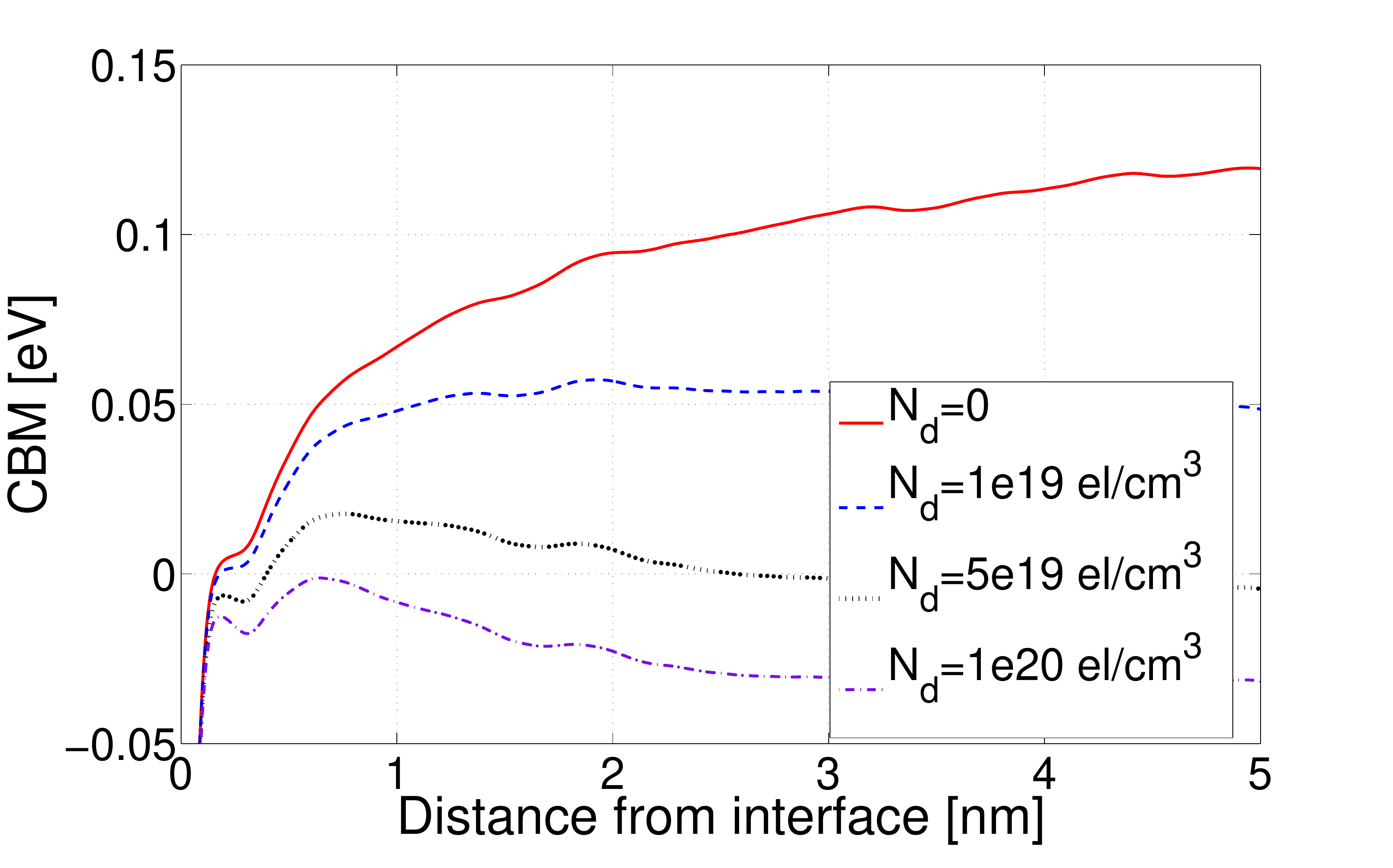}
\includegraphics[width=\columnwidth]{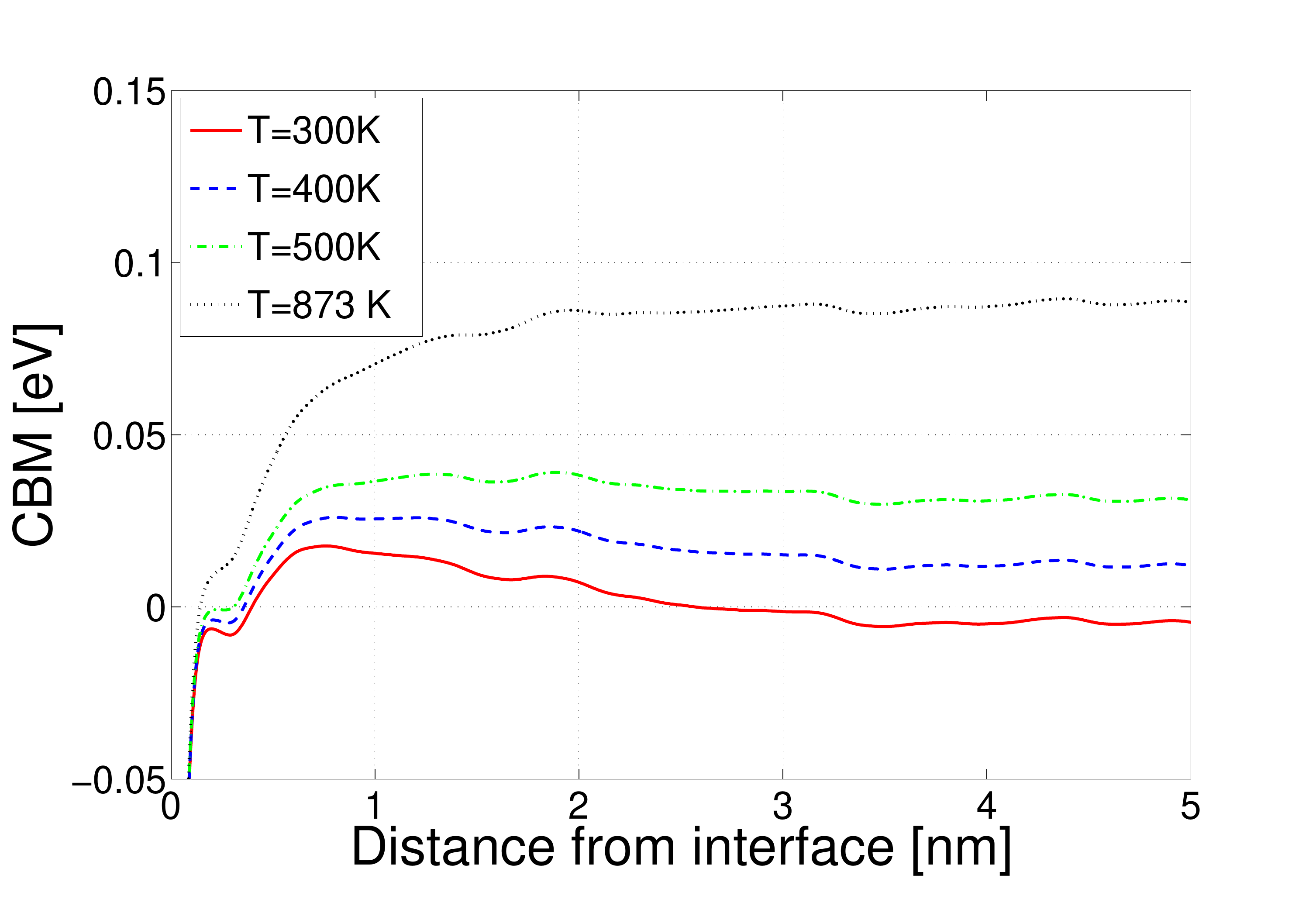}
\end{center}
\caption{a) Band bending at the HH/FH interface calculated using Eq.~\ref{eq_iii}: a) for several n-type doping levels $N_d$ at room temperature. b) for several temperatures at $N_d=5\times10^{19}$ el/cm$^3$.}
\vspace{-0.0cm}
\label{band_bending}
\end{figure}

The results are shown in Fig.~\ref{band_bending} for several doping levels and temperatures (including the optimal operating temperature of $600 ^{\circ}$C). Figure \ref{band_bending}a) shows the CBM profile at room temperature for several n-type doping levels up to $N_d=10^{20}$ el/cm$^3$. We note that the CBM crosses the Fermi level far away from the interface for doping levels higher than $\approx 5\times10^{19}$ el/cm$^3$. At this doping level a Schottky barrier develops near the interface with a small barrier height less than $\approx 50$ meV and a barrier length shorter than $\approx 2$ nm. Figure \ref{band_bending}b) shows that the band bending depends slightly on temperature. This is due to the temperature dependence of the Fermi level w.r.t.~the CBM for a given carrier concentration. We note that as the temperature increases so does the barrier height, raising the question whether the contact resistivity follows a similar but unusual trend.

To calculate the contact resistivity we use an effective mass model \cite{Straton,PRM18} with the band bending potential serving as input. 
Good agreement between the effective mass and full {\it ab initio} approaches has been demonstrated for other semiconductor contacts \cite{Gao}. The effective mass model employs the density of transport modes (DOM) \cite{Lundstrom} in the semiconductor which sets the limit for the minimum  resistivity (achievable in the absence of scattering at the FH/HH interface). The DOM can be parameterized by the effective transport mass $m_{DOM}$ and we use $m_{DOM}=2$ m$_0$ as previously calculated \cite{PRM18}. Together with the Schottky barrier height and length this allows the evaluation of the electron tunneling transmission probability across the barrier. From the transmission function it is then straightforward to obtain the current density and ultimately the contact resistivity \cite{Straton,PRM18}. 

\begin{figure}
\vspace{-0.0cm}
\begin{center}
\includegraphics[width=\columnwidth]{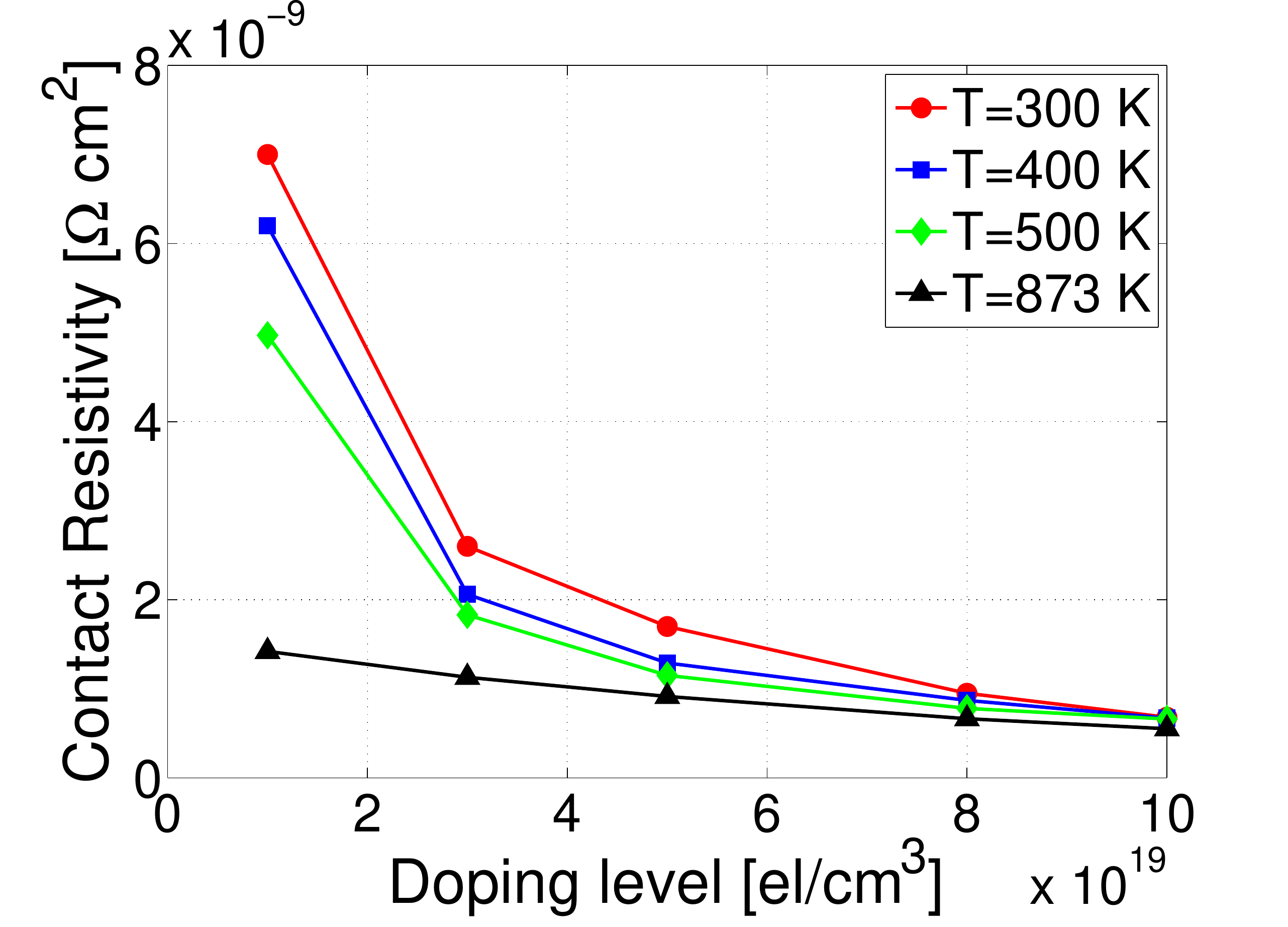}
\end{center}
\caption{Contact resistivity as a function of doping concentration for the n-type HH contacted by FH for several temperatures.}
\vspace{-0.0cm}
\label{contact_resist}
\end{figure}

Figure \ref{contact_resist} shows our main results for the contact resistivity $\rho_C$ of n-doped HH contacted by FH as function of the n-type doping level for several temperatures. The model accounts for several transport channels: tunneling, thermionic, and thermionic field emission. The tunneling mechanism becomes prevalent for the higher doping levels when the Schottky barrier gets smaller and $\rho_C$  approaches the fundamental limit set by DOM. In this regime $\rho_C$ depends negligibly on temperature and the contact is ohmic. This is in agreement well known approaches of doping a semiconductor near the metal contact to minimize the contact resistivity \cite{Taylor,Popovic,Braslau}. 
In the lower doping regime the barrier increases and $\rho_C$  increases together with the weight of the thermionic transport mechanism. Increasing the temperature enhances the probability of thermal excitation over the barrier. Despite the slight increase in barrier height seen in Fig.~\ref{band_bending}b), the enhanced thermal excitation probability leads to the usual global decrease in $\rho_C$  with increasing temperature.  We note that the overall values of the contact resistivity are less than $10^{-8} \Omega$ cm$^2$, two orders of magnitude below what is desired for efficient, compact TEG devices.

In conclusion, we have studied theoretically the structural and electronic properties of the FH/HH interface. The similarity in crystal structure including the good lattice match between FH and HH makes it plausible that the two systems form a coherent interface that could remain stable over a wide range of temperatures. Employing an atomistic approach that combines {\it ab initio} calculations with macroscopic modeling we have calculated  the band bending near the coherent interface and evaluated the contact resistivity. We find excellent contact properties with very low contact resistivity and almost ohmic behavior over a wide range of temperatures and doping levels, suggesting that FH can provide an ideal electrical contact to HH. 

We acknowledge useful discussions with Jeff Sharp (Marlow Industries). This research was developed with funding from the Defense Advanced Research Projects Agency (DARPA) and supported by the DARPA MATRIX program. The views, opinions, and/or findings contained in this article are those of the authors and should not be interpreted as representing the official views or policies of the Department of Defense or the US Government. Sandia National Laboratories is a multimission laboratory managed and operated by National Technology and Engineering Solutions of Sandia, LLC., a wholly owned subsidiary of Honeywell International, Inc., for the US Department of Energy’s National Nuclear Security Administration under Contract No. DE-NA-0003525.

\clearpage
\newpage

\clearpage
\newpage

\begin{thebibliography}{99}
\bibitem{CTE} The linear CTE for the HH HfNiSn is about $10\times 10^{-6}$K$^{-1}$ (or volumetric CTE $\approx30\times 10^{-6}$K$^{-1}$) over the temperature range of interest \cite{Jung}. While we could not find literature data for the HfNi$_2$Sn FH compounds, data is available \cite{Hermet} for the related family of materials TiNi$_2$Sn (FH) and TiNiSn (HH). At room temperature the measured volumetric CTE is $\approx 40\times 10^{-6}$K$^{-1}$ for TiNi$_2$Sn and $\approx 30\times 10^{-6}$K$^{-1}$ for TiNiSn \cite{Hermet}, corresponding to a difference of $\approx30\%$. Because the CTE of the HHs TiNiSn and HfNiSn turn out to be similar, we infer that the CTE of the corresponding FHs TiNi$_2$Sn and HfNi$_2$Sn are also similar. Thus, we assume that the difference between the CTE of HfNiSn and HfNi$_2$Sn is also $\approx30\%$. Using this value, we estimate that the lattice mismatch between HfNi$_2$Sn and HfNiSn will increase by only $0.15\%$ between room temperature and $600 ^{\circ}$.
\bibitem{short_slab} For a shorter HH slab with width $\sim10$ nm, the projection of the electronic wavefunctions in the middle of the slab reveals that $E_F$ tends to align closer to CBM rather than to mid-gap.
\end{thebibliography}
\end{document}